\documentclass{article}

\usepackage{amsmath}
\usepackage{graphicx}
\usepackage{verbatim}
\usepackage{float} 
\usepackage[hidelinks]{hyperref}

\usepackage{natbib}
\bibpunct{(}{)}{;}{a}{}{,} 

\newcommand{\forqs}[0]{\texttt{forqs }}
\newcommand{\codeinput}[1]{\begin{small} \verbatiminput{#1} \end{small}}

\begin{document}

\title{forqs: Forward-in-time Simulation of Recombination, Quantitative Traits, and Selection}
\author{Darren Kessner\,$^{1}$ and John Novembre\,$^{2}$ \\
    $^{1}$Bioinformatics Interdepartmental Program, UCLA \\
    $^{2}$Department of Human Genetics, University of Chicago}

\maketitle

\begin{abstract}

\emph{Summary:} forqs is a forward-in-time simulation of recombination,
quantitative traits, and selection.  It was designed to investigate haplotype
patterns resulting from scenarios where substantial evolutionary change has
taken place in a small number of generations due to recombination and/or
selection on polygenic quantitative traits.

\emph{Availability and Implementation:} \forqs is implemented as a
command-line C++ program.  Source code and binary executables for Linux, OSX,
and Windows are freely available under a permissive BSD license: \\
\href{https://bitbucket.org/dkessner/forqs}{https://bitbucket.org/dkessner/forqs}

\emph{Contact:}
\href{jnovembre@uchicago.edu}{jnovembre@uchicago.edu},
\href{dkessner@ucla.edu}{dkessner@ucla.edu}

\end{abstract}

\section{Introduction}

Simulations have a long history in population genetics, both for verifying
analytical results and for exploring population models that are mathematically
intractable.  Population genetics simulations can be broadly classified as
forward-in-time (e.g. Wright-Fisher) or backward-in-time (e.g. coalescent).
Coalescent simulations (e.g. \texttt{ms} \citep{Hudson2002}, \texttt{MaCS}
\citep{Chen2009}, \texttt{fastsimcoal} \citep{Excoffier2011}) are very
efficient for simulating neutral sequence data because they only need to track
lineages that are ancestral to the sample.  While it is possible to simulate
certain selection scenarios within the coalescent framework
\citep{HudsonKaplan1988, Ewing2010}, one must turn to forward-in-time
simulations in order to model selection in a flexible way.

Many forward-in-time simulators are currently available.  Most of these
simulators use a mutation-centric approach, implemented by tracking the
mutations carried by individuals each generation.  To handle selection, the
majority of these simulators assign selection coefficients to individual
mutations 
(e.g. \texttt{ForwSim} \citep{Padhukasahasram2008} 
\texttt{Fregene} \citep{Chadeau-Hyam2008}
\texttt{GENOMEPOP} \citep{Carvajal-Rodriguez2008} 
\texttt{SFS\_CODE} \citep{Hernandez2008}
\texttt{TreesimJ} \citep{OFallon2010},
\texttt{SLiM} \citep{Messer2013}), 
while a few also include support for quantitative traits 
(e.g.  \texttt{ForSim} \citep{Lambert2008},
\texttt{quantiNemo} \citep{Neuenschwander2008},
\texttt{simuPOP} \citep{Peng2005}).
\cite{Hoban2011} and \cite{Yuan2012} are recent reviews providing
a comprehensive comparison of these and other simulators.

In many scenarios of biological interest, substantial evolutionary change has
taken place in a small number of generations due to recombination and/or
selection on standing variation, rather than mutational input.  For example,
one may be interested in the genome-wide haplotype patterns that emerge from
admixture between historically isolated populations \citep{Wegmann2011},
or from artificial selection on a quantitative trait.  Studying these haplotype
patterns can be difficult with existing forward-in-time simulators, because
information about recombination events is typically not stored due to
computational constraints.  In addition, forward-in-time simulators that
store entire sequences incur a severe trade-off between the size of the
genomic regions and the size of the populations simulated.

Motivated by such examples, we have implemented a new forward-in-time
simulation approach that, instead of tracking single-site variants, tracks
individual haplotype chunks as they recombine over multiple generations.
Further, we have designed the simulator for fast simulation of quantitative
traits under selection.  We have labeled this software \forqs (Forward-in-time
simulation of Recombination, Quantitative Traits, and Selection).  Similar
approaches have been implemented recently by \cite{Haiminen2013} and by
\cite{Aberer2013}, for the simple selection models with per-mutation fitness
effects.  

The haplotype-based design allows for fast simulation of whole genomes, with
very efficient memory usage.  This design preserves information about
recombination events that take place during the simulation, and also allows
for immediate identification of genomic regions where individuals share
identical-by-descent haplotype tracts. Our simulator uses a modular
architecture to allow the user to flexibly specify recombination maps, mutation
rates, demographic models, quantitative traits and fitness functions.  This
modular approach facilitates simulation of complicated scenarios and
investigation of the resulting haplotype patterns, including as an
example, selection for optimal values of a polygenic quantitative trait in
multiple connected populations, where the optimal value may change depending on
the population and generation.  \forqs is currently under active development to
support ongoing projects.

\section{Design and Implementation}

\begin{figure}[!h]
    \centerline{\includegraphics[width=.4\textwidth]{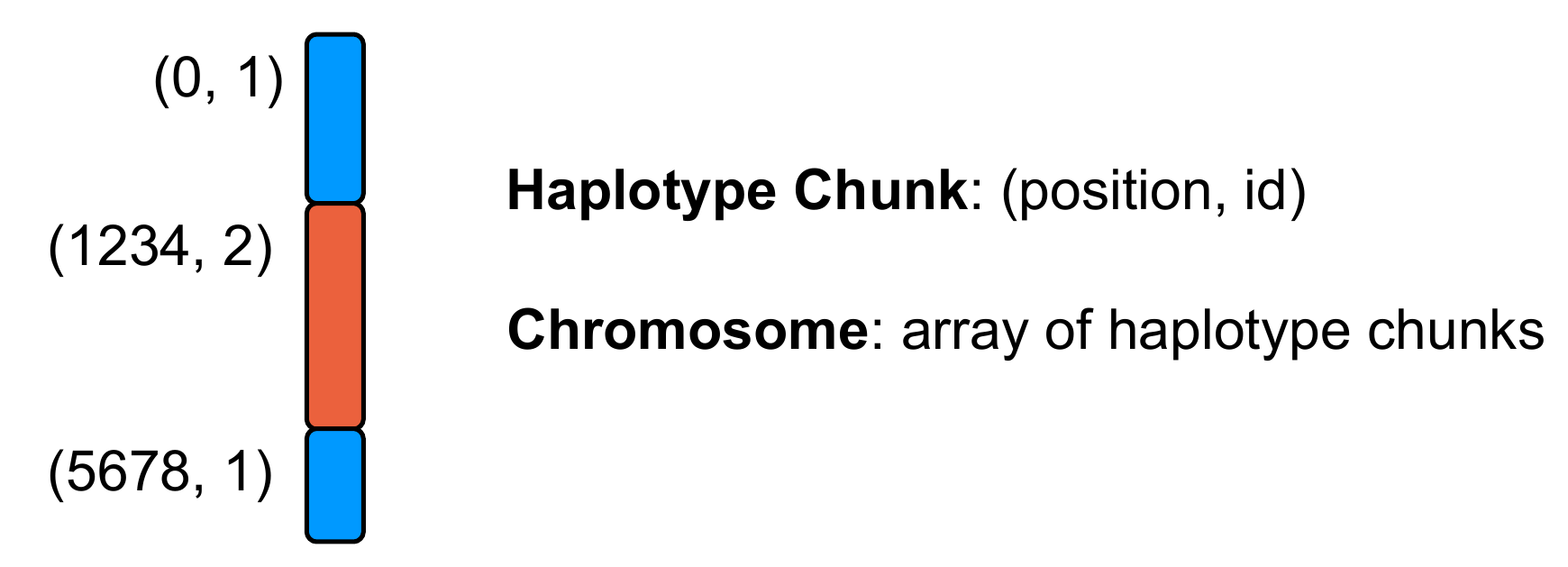}}
    \caption{{\bf \forqs chromosome representation}.  An individual chromosome is  
        represented by a list of haplotype chunks.  Each haplotype chunk is
        represented by two numbers \emph{(position, id)}: the position where it
        begins, and the identifier of the founding haplotype from which it is
        derived.  This cartoon depicts a chromosome with 3 haplotype chunks as
        the result of recombination (double crossover) between two founder
        chromosomes.}
    \label{figure_haplotype_chunk}
\end{figure}

\forqs begins with a set of founding haplotypes representing the individuals in
the initial generation.  By assigning a unique identifier to each founding
haplotype, individual haplotype chunks are tracked as they recombine over
subsequent generations.  For the purposes of simulation, any existing neutral
variation on the haplotype chunks can be ignored.  When simulating selection on
standing variation, only those loci with fitness effects
need to be tracked.

Internally, \forqs represents an individual chromosome as a list of haplotype
chunks (Figure \ref{figure_haplotype_chunk}).  Individuals
are diploid, and carry a user-specified number of chromosome pairs.  To
represent genetic variation at particular loci, \forqs queries functions called
\texttt{VariantIndicators} to obtain the variant values carried by the
founding haplotypes at those loci.

\begin{figure}[th]
    \centerline{\includegraphics[width=.5\textwidth]{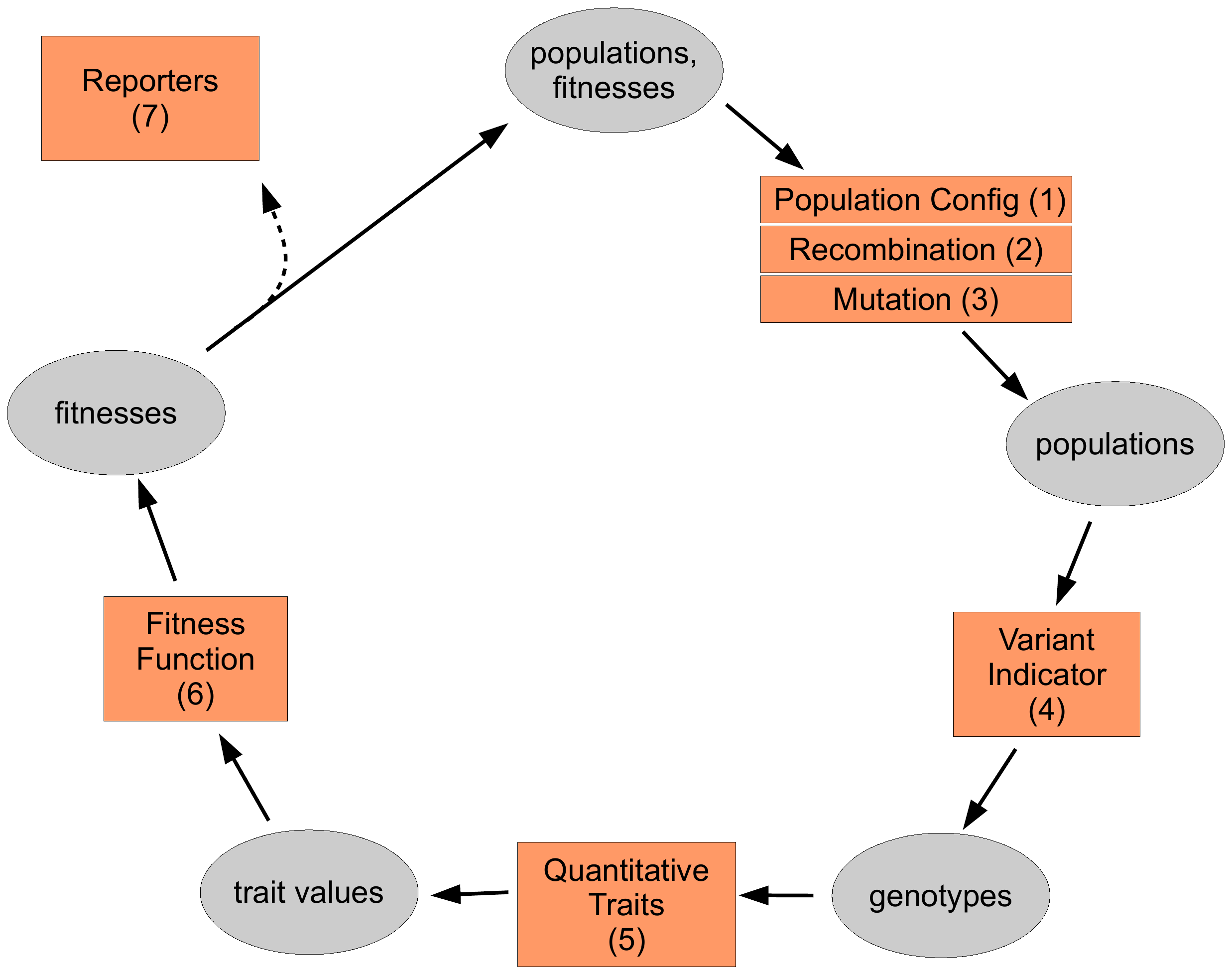}}
    \caption{{\bf \forqs modular design}.  The orange boxes represent places
             where the user can plug in configurable modules.}
    \label{figure_design}
\end{figure}

\vspace{-2ex} 

\forqs uses configurable \emph{modules} that are plugged into the main
simulator to change the behavior of the simulation (Figure
\ref{figure_design}).  \forqs performs the following actions during a single
cycle of the simulation:

\begin{itemize}
    \item[(1-3)] \begin{sloppypar} 
        Generation of new populations: The
        \texttt{Population\-Config\-Generator} module (1) provides the
        simulator with a \emph{population configuration} that specifies how to
        create individuals in the next generation from parents in the previous
        generation, based on user-specified population size and migration rate
        trajectories.  Recombination and mutation are handled by a user-specified
        \texttt{Recombination\-Position\-Generator} module (2) and/or
        \texttt{Mutation\-Generator} module (3), respectively.
        \end{sloppypar}
    \item [(4)] \begin{sloppypar} Genotyping:  individuals are ``genotyped'' at a set of loci,
        using the \texttt{Variant\-Indicator} module to obtain variant (SNP)
        values for each individual.  The list of loci to genotype is determined
        from the quantitative traits and reporters that are specified by the
        user.  \end{sloppypar}
    \item [(5)] Quantitative trait evaluation: for each quantitative trait, a
        trait value is calculated for each individual, based on the
        individual's genotype at loci affecting the trait.  Each quantitative
        trait is specified by a \texttt{Quantitative\-Trait} module.
    \item [(6)] Fitness evaluation:  the \texttt{Fitness\-Function} module
        calculates each individual's fitness based on the individual's 
        quantitative trait values.  
    \item [(7)] Reporting:  each \texttt{Reporter} module updates its output
        files using current information from the populations (which includes
        genotypes, trait values, and fitnesses).
\end{itemize}

Each configurable module in Figure \ref{figure_design} is actually an interface
with multiple implementations.  The user specifies which module implementations
to instantiate in a \forqs configuration file.  In addition to the primary
modules shown in the diagram, there are several building block modules that
provide basic functionality to the primary modules.  For example,
\texttt{Trajectory} modules provide a unified method for specifying values that
change over time, such as population sizes or migration rates.  Similarly,
\texttt{Distribution} modules can be used to specify how to draw particular
random values, for example, QTL positions or allele frequencies.

\begin{sloppypar}
For example, to simulate populations undergoing neutral admixture, the user
would specify a \texttt{Population\-Config\-Generator} module representing a
stepping stone or island model with the desired population size and migration
rate trajectories, and a \texttt{Recombination\-Position\-Generator} with the
desired recombination rate or map, but no quantitative traits.  As another
example, to simulate an artificial selection experiment with truncation
selection on a single quantitative trait, the user would specify the
quantitative trait loci (QTLs) and effect sizes, and choose a
\texttt{Fitness\-Function} module representing truncation selection with the
desired proportion of individuals selected to produce the next generation.
Alternatively, the user could indicate that the QTLs and effect sizes should be
drawn randomly from user-specified distributions.
\end{sloppypar}

The representation of chromosomes as haplotype chunks in \forqs makes very
efficient use of memory, independent of the size of the chromosomes.  However,
memory usage grows linearly with the number of generations simulated, due to
recombination.  On a typical laptop computer, for a population size of 1
million, simulations take $\sim 1.5$ seconds per generation for neutral
simulations and $\sim 3$ seconds per generation with quantitative traits and
selection.  Decreasing the population size allows the simulation of a greater
number of generations in a reasonable amount of time:  for a population size of
10000, it takes $\sim 3$ seconds per 100 generations (without
selection, with a negligible increase with selection).

\forqs has been extensively tested for correctness, both at the level of
individual code units, and in its large-scale behavior in comparison to
theoretical predictions from population genetics and quantitative genetics.
Validation results, as well as tutorials and documentation can be
found in the Supplementary Information.

\section*{Acknowledgements}

The authors would like to thank Alex Platt, Charleston Chiang, Eunjung Han, and
Diego Ortega Del Vecchyo for helpful comments on features, usability, and 
documentation of the software.

This work was supported by the National
Institutes of Health (Training Grant in Genomic Analysis and Interpretation T32
HG002536 for D.K.,  R01 HG007089 for J.N.), the NSF (EF-0928690 for J.N.),
and UCLA (Dissertation Year Fellowship for D.K.).

\clearpage

\pdfbookmark[1]{Supplementary Information}{toc}

\begin{huge}
\begin{center}
    \forqs Supplementary Information
\end{center}
\end{huge}

\begin{center}
    \includegraphics[width=.8\textwidth]{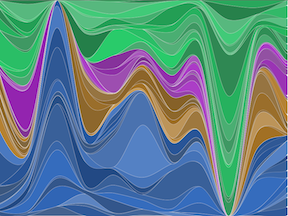}
\end{center}

\forqs is a forward-in-time simulation of recombination,
quantitative traits, and selection.  The \forqs simulator was designed to
investigate haplotype patterns resulting from scenarios where substantial
evolutionary change has taken place in a small number of generations due to
recombination and/or selection on polygenic quantitative traits.  To do this,
\forqs tracks individual haplotype chunks during the course of the simulation;
each chunk carries an identifier specifying the ancestral individual from which
the block is derived.  \forqs is implemented as a command-line C++ program,
using a modular design that gives the user great flexibility in creating
custom simulations.

\forqs is freely available with a permissive BSD license.  Binary executables
can be obtained for Linux, OSX, and Windows from: \\
\url{https://bitbucket.org/dkessner/forqs/downloads}

\section*{About the image}

The image on the first page was created based on haplotype frequency
information reported during a \forqs simulation.  The image shows local
haplotype frequencies (equivalently, local ancestry proportions) along a single
chromosome.  A single population was simulated with selection at two loci (one
near each end of the chromosome).  25\% of individuals (blue) in the initial
generation carried the selected allele at the first locus, and a different 25\%
(green) carried the selected allele at the second locus.  After 100
generations, the two selected variants have fixed in the population.  All
individuals carry one of the blue haplotypes in the region containing the first
locus, as well as one of the green haplotypes in the region containing the
second locus.

\clearpage

\tableofcontents

\newpage

\section{Getting Started}

\forqs binaries (Linux, OSX, and Windows) are distributed in zip file packages
containing the executables, documentation, and example configuration files.
The latest \forqs packages are available here:\\ \indent
\url{https://bitbucket.org/dkessner/forqs/downloads}

\bigskip

\forqs is a command line program that takes a single argument specifying
the configuration file for the simulation:
\begin{small}
\begin{verbatim}
    forqs config_file
\end{verbatim}
\end{small}

The \forqs packages include an \texttt{examples} directory containing several
example configuration files.  After unzipping the \forqs package,
you can run \forqs directly from the package directory:
\begin{small}
\begin{verbatim}
    bin/forqs examples/example_1_locus_selection.txt
\end{verbatim}
\end{small}
\forqs puts all output files in the output directory specified in the
configuration file.  After running this command, you will find a new directory
\path{output_example_1_locus_selection} with the output from this simulation.
For convenience, you can put the \forqs executable somewhere in your
\texttt{PATH}, e.g. \path{~/bin}, so that you don't have to type the path when
running the program.

\begin{figure}[!h]
    \begin{center}
        \includegraphics[width=1\textwidth]{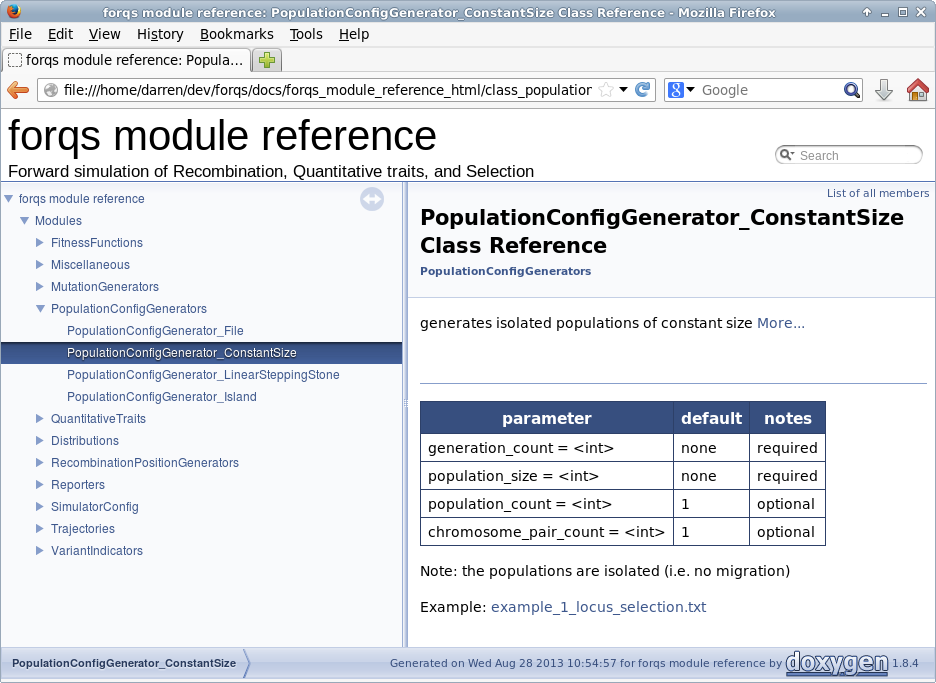}
    \end{center}
    \caption{{\bf \forqs Module Reference}.}
    \label{figure_forqs_module_reference}
\end{figure}

In addition to the document you are now reading, \forqs packages include the
\forqs Module Reference (\path{docs/forqs_module_reference.html}, Figure
\ref{figure_forqs_module_reference}), which contains details about each module,
including parameter names, usage, and links to examples.

\clearpage

\section{Tutorial Introduction}

In this section we introduce \forqs configuration files, starting with a
minimal example and gradually adding modules to add complexity to the
simulation.  All example configuration files can be found in the \texttt{examples}
directory of the \forqs package.

\subsection{A minimal example}

We start with a minimal example configuration file:
\codeinput{tutorial_0_minimal.txt}

There are two modules specified in this configuration file.  The first module
is \texttt{PopulationConfigGenerator\_ConstantSize}, which we have assigned the
object id \texttt{pcg}.  The object id, which can be any text string, allows
other objects to reference this object.  We have set two parameters for this
module, which specify the population size (\texttt{population\_size = 10}) and 
the number of generations to simulate (\texttt{generation\_count = 3}).

The second module is \texttt{SimulatorConfig}, the top-level module.   This
module must be specified last in the configuration file, because it has
references to the primary modules (the general rule is: objects must be
specified before they can be referenced by other objects).  In this case, we
are telling \texttt{SimulatorConfig} to use object \texttt{pcg} as our
\texttt{PopulationConfigGenerator}, which is the only primary module that is
required to be specified.  The other primary modules have trivial
defaults, and we leave them unspecified.  In addition to the primary module
references, \texttt{SimulatorConfig} also contains global simulation
parameters.  In this case, we have specified the output directory to be
\path{output_tutorial_0_minimal}.

The simulator queries the \texttt{PopulationConfigGenerator} each generation to
obtain a population configuration, which contains the information necessary to
create the next generation from the previous one.  In this case, the population
configuration is the same each generation --- it tells the simulator to create a
population of 10 individuals with parents drawn from the same population in
the previous generation.

When you run \forqs on this example (\path{tutorial_0_minimal.txt}),
you will see screen output showing that the simulation ran for 3 generations.
You will also see the new output directory \path{output_tutorial_0_minimal}.
If you look in this directory, you'll see a single file:
\path{forqs.simconfig.txt}.  This file contains the configuration used for the
simulation, including unspecified default parameters.  However, there are no
other output files in the directory, so we don't actually know what happened
during the simulation.  To get more information, we need to specify Reporters to
report output --- we show how to do this in the next section.

\subsection{Recombination and reporting output}

In this example, we add modules to include recombination and report output:
\codeinput{tutorial_1_recombination_reporter.txt}

\begin{sloppypar}
When creating offspring chromosomes from parental chromosomes, the simulator
uses the \texttt{RecombinationPositionGenerator} module to generate lists of
recombination positions.  In this case, we are using
\texttt{RecombinationPositionGenerator\_Uniform}, which chooses the positions
uniformly at random along the chromosome.  Note the use of the compound
parameter \texttt{chromosome\_length\_rate} that specifies both the length of the
chromosome and the recombination rate.
\end{sloppypar}

\texttt{Reporter} modules are used by \forqs to output information about the
simulation.  The user can specify an arbitrary number of \texttt{Reporters},
depending on what output is desired.  \texttt{Reporter\_Population} outputs
\forqs population files, where each individual is represented as a list of
chromosomes (one chromosome per line) and each chromosome is represented as a
list of haplotype chunks.  By default, population files will be produced only
for the final generation --- setting \texttt{update\_step = n} will tell the
\texttt{Reporter\_Population} to output population files every \texttt{n}
generations.  You can see this by uncommenting (removing the '\#' character)
the \texttt{update\_step} line in the configuration file and re-running the
simulation.

The population files contain the raw information about the simulated
individuals, and are not meant to be analyzed directly.  Instead, they can be
used to propagate existing neutral variation from the founding haplotypes of
the initial generation to the mosaic haplotypes of the final generation
(see Section \ref{section_background_variation}).

However, we will use the population files to illustrate \texttt{forqs}' internal 
representation of chromosomes.  The first chromosome in the population file
from the final generation is:
\begin{small}
\begin{verbatim}
+ { (0,12) (704958,7) }
\end{verbatim}
\end{small}
This means that the first chromosome of the first individual is made up of two
haplotype chunks: the first chunk (0,12) starts at position 0 and comes from
founder individual 12; the second chunk (704958,7) starts at position 704958
and comes from founder individual 7.
Now comment out the following line in \texttt{SimulatorConfig}:
\begin{small}
\begin{verbatim}
    recombination_position_generator = rpg
\end{verbatim}
\end{small}
which 'un-plugs' the \texttt{RecombinationPositionGenerator} from the simulator.
Now re-run the simulation, specifying a different output directory on the command line:
\begin{small}
\begin{verbatim}
    forqs tutorial_1_recombination_reporter.txt output_directory=out2
\end{verbatim}
\end{small}
When you look at the resulting final population file, you will see that all
chromosomes consist of a single chunk, as you would expect with no recombination.

\subsection{Wright-Fisher simulation}

Our next example is a simple Wright-Fisher simulation where we track the
allele frequency at a single locus under neutral drift:

\codeinput{tutorial_2_wright_fisher.txt}

In this example, the population size is 100, the initial allele frequency is
.5, and the simulation runs for 200 generations.

The \texttt{Locus} module defines a single site --- in this case it is position
100000 on chromosome 1.  The \texttt{Locus} has id \texttt{my\_locus}, and this
id is used by two other modules to refer to this locus.

As we saw in the previous example, \forqs represents chromosomes as lists of
haplotype chunks, with no information about nucleotides or variant (SNP) values
at any position.  (To be concrete, we think of a variant value as being 0 or 1,
but \forqs variant values can be anything in the range 0--255, so we are not
restricted to bi-allelic SNPs).  However, if we give \forqs the variant values
for the founding haplotypes, it can calculate variant values for any mosaic
chromosome.  This information is represented by a \texttt{VariantIndicator},
which is a function that tells \forqs which founding chromosomes contain which
variants.

In our case, we specify a \texttt{VariantIndicator\_SingleLocusHardyWeinberg},
which assigns variants to individuals in Hardy-Weinberg proportions.  Because
we have specified an initial allele frequency of .5, this means that of the
founding individuals, 25\% will be homozygote 0, 25\% will be homozygote 1, and
50\% will be heterozygotes.

We also want to track the allele frequency at this locus, so we specify a
\texttt{Reporter\_AlleleFrequencies}, which reports the allele frequency
at each generation.  You will find the output in the file 
\path{allele_frequencies_chr1_pos100000.txt}.

\begin{sloppypar}
You may have noticed the commented line with the parameter \texttt{population\_count = 10}.
By uncommenting this line, you will tell \forqs to simulate 10 populations.  
\texttt{PopulationConfigGenerator\_ConstantSize} does not allow migration between 
populations, so this effectively gives 10 independent Wright-Fisher simulations.  The
resulting allele frequency trajectories can all be found in the same output file,
with one column per population.  As expected, you will see that the alternate allele
(1) has been fixed or lost in some populations, and is still polymorphic
in others.
\end{sloppypar}

\subsection{Selection}

In this example, we add selection to our Wright-Fisher simulation.

\codeinput{tutorial_3_selection.txt}

For this simulation, we add a quantitative trait that is determined by the
genotype at a single locus.  In fact, in our case, the quantitative trait
\emph{is} fitness (\texttt{QuantitativeTrait\_SingleLocusFitness}), and we use
the identity function for our fitness function
(\texttt{FitnessFunction\_Identity}).  The parameters \texttt{w0, w1, w2} are
used to specify the fitness for individuals with genotype 0, 1, and 2,
respectively.  In general, a quantitative trait can depend on multiple loci on
multiple chromosomes, and fitness can depend on multiple quantitative traits.

\begin{figure}[!h]
    \begin{center}
        \includegraphics[width=.60\textwidth]{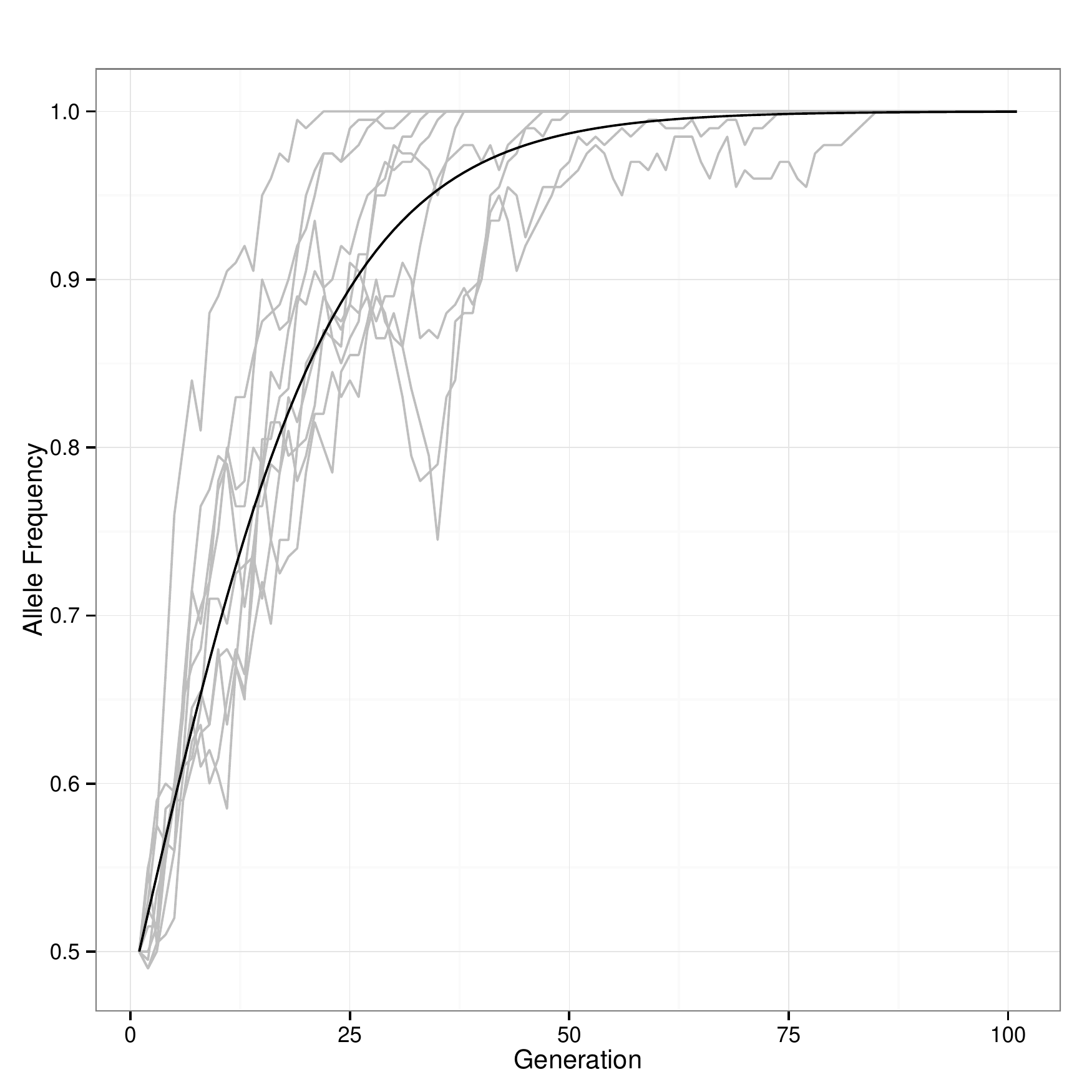}
    \end{center}
    \caption{{\bf Allele frequency trajectories from selection simulation}.}
    \label{figure_allele_frequency_trajectories}
\end{figure}

We also add some useful \texttt{Reporters} that output the mean fitnesses of the
populations as well as the deterministic trajectories expected in the limit of
infinite population size.  These output files can be easily read into a
plotting application (e.g. \texttt{R}) to compare the random trajectories with
the deterministic trajectories (Figure \ref{figure_allele_frequency_trajectories}).

\subsection{Trajectories}

This section shows how to use \texttt{Trajectory} modules to specify population
sizes that change over time.  In general, \texttt{Trajectory} modules can be used
to specify any numeric value that varies by population and generation -- other examples
include mutation rates and migration rates.

\begin{sloppypar}
In the first example (\path{tutorial_4a_trajectories.txt}), there are two populations
following the same population size trajectory: constant size 100 until generation 2,
followed by a linear increase in population size until it reaches 400 at generation 5,
after which the population size stays at 400.  This is accomplished by specifying each
piece (\texttt{Trajectory\_Constant}, \texttt{Trajectory\_Linear}, \texttt{Trajectory\_Constant}) 
and composing the pieces together with \texttt{Trajectory\_GenerationComposite}.  
\end{sloppypar}

\codeinput{tutorial_4a_trajectories.txt}

\begin{sloppypar}
Note that \texttt{PopulationConfigGenerator\_LinearSteppingStone} has a
\texttt{population\_size} parameter whose value must be the id of a
\texttt{Trajectory} module (rather than a numeric value).  Also,
unless otherwise specified, each \texttt{Trajectory} gives the same numeric
value for each population in a given generation.  You can see the raw
population configuration generated by including the parameter
\texttt{write\_popconfig = 1}, which tells \forqs to write out the population
configs in a file \path{forqs.popconfig.txt}.  After running this example, you
can examine \path{forqs.popconfig.txt} to see the population sizes.
\end{sloppypar}

In the next example (\path{tutorial_4b_trajectories.txt}), population 1 follows
the same population size trajectory as in the first example, while population 2
experiences exponential growth, doubling each generation (with initial
population size 1000).  In order to specify this scenario, each population size
trajectory is specified separately (\texttt{popsize\_pop1} and
\texttt{popsize\_pop2} in this example), and the two are composed with
\texttt{Trajectory\_PopulationComposite}.  When
\texttt{Trajectory\_PopulationComposite} is queried for a value for a
population, it returns the value obtained from the appropriate sub-trajectory.

\codeinput{tutorial_4b_trajectories.txt}

\subsection{Quantitative traits}

In this section, we present an example of specifying a quantitative trait
by specifying quantitative trait loci (QTLs) and their effect sizes.

\codeinput{tutorial_5_qtl.txt}

In this example, we specify a single population (size 100) where individuals
have 3 chromosomes.  We specify 3 loci, each on a different chromosome.

\texttt{VariantIndicator\_Random} is used to assign variants to the initial haplotypes 
randomly, according to the specified allele frequencies -- in this case, the
variant allele frequency is .5 for each locus.  The \texttt{'*'} indicates
that all populations have the same allele frequencies -- because there is only one
population, we could have used \texttt{'1'} in place of \texttt{'*'}.  The
parameter \texttt{write\_vi = 1} in \texttt{SimulatorConfig} tells \forqs to
write the file \path{forqs.vi.txt} containing lists of the initial haplotypes carrying
the variant allele at each locus.

\texttt{QuantitativeTrait\_IndependentLoci} specifies a quantitative trait where
the effects of variants at different loci are independent of each other, i.e. there
are no epistatic effects.  Each QTL is listed with its per-genotype effects -- in 
this case, individuals with genotype 0, 1, or 2 at \texttt{locus1} will have
0, .1, or .2 added to their trait value, respectively.  Alternatively, QTL effect
sizes and dominance values can be chosen randomly according to specified
distributions.  For an example of this, see \path{example_qtl_random.txt}
which also demonstrates the use of \texttt{LocusList\_Random} to pick the loci
randomly as well.

\begin{sloppypar}
In this example, \texttt{FitnessFunction\_TruncationSelection} assigns fitness
values of 1 to individuals in the upper half of the distribution, and 0 to
individuals in the lower half (\texttt{proportion\_selected = .5}).
\texttt{Reporter\_AlleleFrequencies} uses the parameter
\texttt{quantitative\_trait = qt}, which tells it to report allele frequencies
for all QTLs associated with the trait.  \texttt{Reporter\_TraitValues} reports
mean trait values for the population each generation; the parameter
\texttt{write\_full = 1} tells it to also report individual trait values for
each generation.
\end{sloppypar}

\subsection{Reporting haplotype frequencies}

This example shows the use of \texttt{Reporter\_HaplotypeFrequencies} to
report local haplotype frequencies in the simulated populations.  

\begin{sloppypar}
In this example, there is a single locus that confers an additive fitness
advantage to the individuals carrying the selected variant.  The
\texttt{VariantIndicator} specifies which haplotype ids actually carry the
variant:  in this case, \texttt{VariantIndicator\_IDRange} assigns the value
\texttt{1} to the first 1000 haplotype ids, i.e. the first 500 individuals are
homozygous for the selected variant.
\end{sloppypar}

\texttt{Reporter\_HaplotypeFrequencies} uses a \texttt{HaplotypeGrouping} module
that specifies how haplotype ids should be grouped.  The grouping can be as
fine-grained as a single haplotype id per group, but for simplicity we have
specified just two groups:  those individuals who carry the selected variant,
and those who don't.  \texttt{Reporter\_HaplotypeFrequencies} then reports
local proportions of the two groups in the population.

We have specified a relatively high recombination rate (3 crossovers per
meiosis on average) and strong selection (.1 selection coefficient) in order to
illustrate the change in local haplotype frequencies in just a few generations.
After running the example, the output file
\path{haplotype_frequencies_chr1_final_pop1.txt} shows that haplotypes carrying
the selected variant have risen to high frequency near the selected locus, but
not as much in regions farther away, due to recombination.  The parameter
\texttt{update\_step = k} will result in reporting of local haplotype
frequencies every \texttt{k} generations (default is to report only the final
generation).

\codeinput{tutorial_6_haplotype_frequencies.txt}

\newpage

\section{Modules}
\label{section_modules}

\begin{sloppypar}
In this section, we give an overview of \forqs modules.  For information about
specific \forqs modules, including parameter names, usage, and links to
examples, please refer to the \forqs Module Reference
(\texttt{forqs\_module\_reference.html}).
\end{sloppypar}

\subsection{Module types and interfaces}

\forqs has a single \emph{top-level module} called \texttt{SimulatorConfig}.
There are seven \emph{primary modules}, corresponding to the orange boxes
Figure \ref{figure_design}, which represent the main places where \forqs
can be customized.  There are also several \emph{building block modules},
which are used by the primary modules.

Multiple modules may share a common interface, as indicated by their names.
For example, \texttt{Reporter\_Timer} and \texttt{Reporter\_AlleleFrequencies} both implement the
\texttt{Reporter} interface.  Modules with the same interface can be used
interchangeably by the simulator.  

In addition, specific modules can be instantiated multiple times (with
different object ids and parameters) -- for example,
\texttt{Reporter\_AlleleFrequencies} can be used to report allele frequencies at several
different loci.

By mixing and matching different modules, the user has a great deal of
flexibility in creating customized simulation scenarios.  In addition, \forqs
can easily be extended with new functionality by adding new modules that implement
existing module interfaces.

\subsection{Module specification and instantiation}

Each \forqs module specified in the configuration file corresponds to an object
that is instantiated by the simulator before the simulation starts.  The first
line of a module specification contains the module name and object id.  Each
subsequent line specifies a parameter of the module, formatted as a name-value
pair: \emph{name = value}.  The value is read as a string, but may be
interpreted as an integer, floating point, string, or list, depending on the
parameter.   Each module determines how its parameter values are interpreted,
and interpretation errors are reported to the user.  In some modules, multiple
parameter values may be specified with the same name, in which case each value
specified is appended to a list.  A blank line ends the specification for that
module.

As an example, here is the specification of a \texttt{Locus} module, which 
represents a single nucleotide position:
\begin{small}
\begin{verbatim}
Locus my_locus
    chromosome = 1
    position = 500000
\end{verbatim}
\end{small}
The first line specifies a \texttt{Locus} object with id
\texttt{my\_locus}.  The two parameters specify which
chromosome (1) and the position on the chromosome (500000).

After parsing this module specification, the simulator instantiates a
\texttt{Locus} object, and puts the object in a registry under the name
\texttt{my\_locus}.  Subsequently instantiated objects may need a reference to
this object, which can be obtained by looking up the object by name in the
registry.  Because modules are instantiated in the order they are specified in
the configuration file, it is important to specify an object before it is
referenced by other objects (otherwise \forqs will produce an error message
that it couldn't find the referenced object in the registry).

Object references are specified as string-valued parameters, where the value
contains the object id of the referenced object.  For example, the following is
the specification of a \texttt{Reporter} module that holds a reference to our
example \texttt{Locus} object \texttt{my\_locus}:
\begin{small}
\begin{verbatim}
Reporter_AlleleFrequencies reporter_allele_frequencies
    locus = my_locus
\end{verbatim}
\end{small}

One final note on a parameter naming convention used in \texttt{forqs} -- for
parameters where the value is a list of sub-parameters, the parameter name should be a
concatenation of the sub-parameters.  This convention allows easier reading
and editing of configuration files without the need to consult documentation.  
For example, \texttt{Trajectory\_GenerationComposite} represents a list of \texttt{Trajectory}
modules, to be followed piecewise at specified generations.  This module is 
parametrized by a list of pairs (generation, trajectory),
using the parameter name \texttt{generation\_trajectory}:
\begin{small}
\begin{verbatim}
Trajectory_GenerationComposite id_migration_rate
    generation_trajectory = 0 id_migration_rate_0
    generation_trajectory = 5 id_migration_rate_1
\end{verbatim}
\end{small}
In this case, the module configuration can be interpreted as follows: at
generation 0 use trajectory \texttt{id\_migration\_rate\_0}, and then at
generation 5 start using trajectory \texttt{id\_migration\_rate\_1}.

\subsection{Top-level module: SimulatorConfig}

\texttt{SimulatorConfig} is the top-level module containing:
\begin{itemize}
    \item global simulation parameters
    \item references to the primary modules
\end{itemize}
Because \texttt{SimulatorConfig} has references to the primary modules, it must
be specified last in the configuration file.

\subsubsection{Global simulation parameters}

The main global simulation parameters are:
\begin{itemize}
    \item \texttt{output\_directory}: \forqs will create this directory and
        place all output files here
    \item \texttt{seed}: seed for the random number generator
\end{itemize}

Command line parameters can also be specified on the command line as
\emph{name=value} with no whitespace, as shown in this example:
\begin{small}
\begin{verbatim}
    forqs config.txt output_directory=mydir seed=12345
\end{verbatim}
\end{small}
If a parameter is specified on both the command line and in the configuration
file, the command line takes precedence.  

If no random seed is specified, \forqs will first look in the current working
directory for the file \path{forqs.seed} containing a previously generated
seed.  If no seed is found, \forqs will generate a new seed from the system
time.  At the end of each run, \forqs generates a random seed and writes it to
\path{forqs.seed}.  To summarize, \forqs looks for the seed in the following
places, in order of preference:
\begin{enumerate}
    \item command line
    \item configuration file
    \item \path{forqs.seed}
    \item system time
\end{enumerate}

\subsubsection{References to primary modules}

Of the primary modules, only \texttt{PopulationConfigGenerator} is required
to be specified, because it contains necessary information about population
sizes and the number of generations.  The rest of the modules default to
trivial implementations. 
The primary modules specified in \texttt{SimulatorConfig} are:
\begin{enumerate}
    \item \texttt{PopulationConfigGenerator} 
        \begin{itemize}
            \item required -- no default
        \end{itemize}
    \item \texttt{RecombinationPositionGenerator} 
        \begin{itemize}
            \item default: \texttt{Recombination\-Position\-Generator\-\_Trivial} 
            (no recombination -- whole chromosomes are transmitted, chromosome pairs
            segregate independently)
        \end{itemize}
    \item \texttt{MutationGenerator}
        \begin{itemize}
            \item default: none (no mutation)
        \end{itemize}
    \item \texttt{VariantIndicator}
        \begin{itemize}
            \item default: \texttt{VariantIndicator\_Trivial} (always returns 0)
        \end{itemize}
    \item \texttt{QuantitativeTrait}
        \begin{itemize}
            \item default: none (no quantitative traits)
            \item multiple \texttt{QuantitativeTraits} may be defined
        \end{itemize}
    \item \texttt{FitnessFunction}
        \begin{itemize}
            \item default: \texttt{FitnessFunction\_Trivial} (always returns 1 -- all individuals have equal fitness)
        \end{itemize}
    \item \texttt{Reporter}
        \begin{itemize}
            \item default: none (no reporters)
            \item multiple \texttt{Reporters} may be defined
        \end{itemize}
\end{enumerate}

\subsection{Primary modules}

\subsubsection{PopulationConfigGenerator}

For the reproduction/transmission step, in addition to information about the current
populations, the simulator needs to know the \emph{population configuration}
for the next generation, which includes:
\begin{itemize} 
    \item number and sizes of the populations in the next generation 
    \item mating distribution, which describes how parents are chosen from the
          current generation to create offspring in the next generation 
\end{itemize}
The \texttt{Population\-Config\-Generator} module provides an interface for the simulator
to obtain a population configuration for each generation.

For the initial generation, individuals are assigned haplotype ids sequentially
starting at 0, with two ids assigned per individual (one for the maternal
chromosomes, and one for the paternal chromosomes).  Id offsets can be used
to make it easier to distinguish between populations -- for example, individuals
from population 1 may be assigned ids starting at 0, and individuals from
population 2 may be assigned ids starting at 1000000.

\texttt{Population\-Config\-Generator\-\_File} allows specification of a population
configuration for each generation, giving the user precise control over the
demographic histories of the simulated populations.

Alternatively, \forqs provides higher level \texttt{PopulationConfigGenerators}
that allow the user to specify time-dependent population size and migration
rate trajectories (e.g.
\texttt{Population\-Config\-Generator\_LinearStepping\-Stone}).  When a
migration rate is specified from a source population to a destination
population, it is interpreted to be the probability that a new child in the
destination population has parents in the source population.  Equivalently, it
is the expected proportion of individuals in the next generation of the
destination population whose parents were in the source population.

For debugging purposes, \forqs optionally outputs the file \path{forqs.popconfig.txt} 
with the population configuration that was used for each generation during the simulation.
This option can be selected by setting the \texttt{write\_popconfig} parameter in
\texttt{SimulatorConfig}:
\begin{small}
\begin{verbatim}
    write_popconfig = 1
\end{verbatim}
\end{small}

\subsubsection{MutationGenerator}

The user may specify a \texttt{MutationGenerator} module to generate random
mutations during the simulation.  \texttt{MutationGenerator} provides a single
interface through which the simulator obtains a list of new mutations for a
particular generation.  However, \texttt{MutationGenerator} implementations may
differ in how they generate this list of mutations.  For example,
\texttt{Mutation\-Generator\_Single\-Locus} generates mutations only at a
single site, while \texttt{MutationGenerator\_Regions} generates mutations in
multiple regions with different (possibly time-dependent) mutation rates.  In
any case, the user will most likely want to also specify
\texttt{Reporter\_Regions} to report the final mutated sequences at the end of
the simulation.

\forqs current mutation implementation creates a new haplotype id for each new
mutation, storing it in a haplotype ancestry tree (so that information about
the ancestral haplotype can be preserved).  This results in memory usage that
increases with each generation, which may cause performance degradation in
simulations involving a high total mutation rate for a large number of
generations.  (Here \emph{total mutation rate} means $\theta L$, where $\theta
= 4N\mu$ is the population-scaled per-site mutation rate and $L$ is the total
length of the genomic region where new mutations are being generated).

\subsubsection{RecombinationPositionGenerator}

The \texttt{RecombinationPositionGenerator} module provides an interface for
the simulator to obtain a list of recombination positions -- the position list
is used to create an offspring chromosome from a pair of parental chromosomes.

\texttt{Recombination\-Position\-Generator\-\_Recombination\-Map} generates
recombination positions according to previously estimated recombination maps
(e.g. from HapMap project).


\subsubsection{VariantIndicator}
\label{subsection_variant_indicator}

Internally, \forqs represents each individual chromosome as a list of haplotype
chunk ids, with no information about particular variants (e.g. SNP 0/1 values)
carried on that chromosome.  In order to obtain an individual's genotype at a
locus, the simulator must use a \texttt{VariantIndicator}, which is essentially
a function that maps:
\begin{align*}
    (\emph{locus}, \emph{haplotype id}) \mapsto \emph{variant value}
\end{align*}

In general, a \texttt{VariantIndicator} needs to specify variant values only at
selected loci, because neutral loci have no effect on the simulation.
Ancestral neutral variation that is present on the haplotypes of the founders
(individuals in the initial generation) can be propagated to individuals at the
end of the simulation using the \texttt{forqs\_map\_ms} tool included in the
\forqs package.

\subsubsection{QuantitativeTrait}

A \texttt{QuantitativeTrait} module represents a single quantitative trait.
The module encapsulates the information needed to calculate the trait value for
each individual based on the individual's genotypes at a list of user-specified
loci.  For example, \texttt{QuantitativeTrait\_IndependentLoci} allows the user
to specify loci and effect sizes, and the trait value for each individual is
calculated by combining that individual's locus-specific effects additively
(with user-specified environmental variance).

\subsubsection{FitnessFunction}

A \texttt{FitnessFunction} module represents the function used to calculate an
individual's fitness based on the individual's trait values.  In simple cases,
the quantitative trait may actually be fitness, in which case the user may
simply use \texttt{FitnessFunction\_Identity}.  In more complicated cases,
fitness may depend non-trivially on one or more quantitative traits.

\subsubsection{Reporter}

\texttt{Reporter} modules are used to report information during the simulation.
For example, \texttt{Reporter\_AlleleFrequencies} reports the allele frequency
of a user-specified locus at each generation.  The full list of available
\texttt{Reporters} can be found in the \forqs Module Reference.

\subsection{Building block modules}

\subsubsection{Locus and LocusList}

The \texttt{Locus} module represents a single site on a chromosome, specified
by the chromosome pair index and position on the chromosome:

\begin{small}
\begin{verbatim}
Locus my_locus
    chromosome_pair_index = 0
    position = 500000
\end{verbatim}
\end{small}

Note that the chromosome pairs use 0-based indexing, so \texttt{chromosome\_pair\_index = 0}
refers to the first chromosome pair.

\texttt{LocusList} modules are used to define a list of loci.  The list can be specified
in two ways -- in the first, loci are specified by chromosome number and position:
\begin{small}
\begin{verbatim}
LocusList locus_list
    chromosome:position = 1 1000123
    chromosome:position = 2 2000234
    [...]
\end{verbatim}
\end{small}

Alternatively, the loci can be references to previously specified \texttt{Locus} objects:
\begin{small}
\begin{verbatim}
LocusList locus_list_2
    loci = my_locus_1 my_locus_2 [...]
\end{verbatim}
\end{small}

Also, a random list of loci can be generated with \texttt{LocusList\_Random}:
\begin{small}
\begin{verbatim}
LocusList_Random locus_list_random
    locus_count = 10
\end{verbatim}
\end{small}

\subsubsection{Trajectory}

In simple situations, constant parameter values can be used to specify an
aspect of the simulation.  For example, it is common in population genetics to
simulate models where population sizes are constant.  In more complicated
scenarios, parameter values (e.g. population sizes, migration rates, mutation
rates, optimal quantitative trait values) may vary in space (population) or in
time (generation).

\forqs uses \texttt{Trajectory} modules to describe such varying quantities.
In essence, a \texttt{Trajectory} represents a function:
\begin{align*}
    (\emph{population index}, \emph{generation index}) \mapsto \emph{value}
\end{align*}

Trajectories can be built by the user by starting with simple building
blocks (e.g. constant, linear, polynomial, exponential functions) and then
composed either by population or by generation.  Once the \texttt{Trajectory}
has been defined, it can be referenced by name to be used as the parameter of
another module.  Note that if a module expects a particular parameter to be a
\texttt{Trajectory}, it is an error to pass a constant integer or floating
point number as the parameter value.  In this case, if the user really wants a
constant parameter value, \texttt{Trajectory\_Constant} should be specified.

\subsubsection{Distribution}

\texttt{Distribution} modules represent probability distributions.
These modules are used to model, for example, effect sizes for QTLs or
environmental variance for a quantitative trait.

\texttt{Distribution\_Constant} can be used in situations where a module
requires a reference to a \texttt{Distribution} but a constant value is desired.

\newpage

\section{Simulating background variation}
\label{section_background_variation}

Forward-in-time simulators often use a burn-in period to allow neutral
variation to reach mutation-drift equilibrium.  An alternative strategy,
adopted by \texttt{forqs}, is to use an existing program (such as Dick Hudson's
\texttt{ms}) to generate neutral variation for the founders in the initial
population.  Chromosomes in the final populations are mosaics of the founder
chromosomes, so neutral variation can be propagated to the mosaic chromosomes to
generate sequence (or SNP) data.  The tool \texttt{forqs\_map\_ms} is included
in \forqs packages for this purpose.  In this section we discuss the propagation
procedure using two examples.  

\subsection{Example 1:  No new mutations}

The first example is a simple case where there are no new mutations introduced
during the \forqs simulation.  This case includes many scenarios of interest
where the variation in the population has been generated over a small number of
generations, primarily by recombination and recent admixture between
historically isolated (or inbred) populations.

In this case, the propagation of neutral variation is straight-forward:
sequence variants on founder haplotypes are mapped onto the mosaic chromosomes
of the final population.  However, there are many details about the mapping
that must be specified by the user in a mapping configuration file.  \forqs
produces final populations of individuals that may carry multiple pairs of
chromosomes, with absolute positions specified by integers.  \texttt{ms}
outputs single chromosomes, with relative positions specified by floating point
numbers.  The mapping configuration thus needs to specify an integer range of
positions on a particular chromosome that correspond to the \texttt{ms}
relative positions in the range [0,1].  In addition, the user must specify how
\forqs haplotype ids correspond to the \texttt{ms}-format variant sequences.

In the \texttt{examples} directory, the following 4 files can be found:
\begin{itemize}
    \item \texttt{population\_example.txt} (\forqs population data)
    \item \texttt{ms\_test\_data\_1.txt, ms\_test\_data\_2.txt} (\texttt{ms}-format files)
    \item \texttt{ms\_map\_config.txt} (mapping configuration file)
\end{itemize}

Perform the mapping using the following command:
\begin{small}
\begin{verbatim}
forqs_map_ms population_example.txt ms_map_config.txt > output.ms
\end{verbatim}
\end{small}

The resulting sequences in \texttt{output.ms} are mosaics of the sequences in
the input \texttt{ms}-format files.  For example, the first chromosome of the
first individual is a mosaic of haplotypes 0, 1, 2, and 3, corresponding to the
sequences 'aaaaaaaaa', 'bbbbbbbbb', 'ccccccccc', and 'ddddddddd', respectively.
The output sequence for this chromosome is 'aabccccdd'.  Note that alphabetic
letters are used in these example sequences for illustration only -- real
\texttt{ms} output consists of 0's and 1's.

\subsection{Example 2:  Including new mutations}

In scenarios where it is important to include new mutations in addition to
ancestral neutral variation, the mapping procedure has one additional
complication.  When simulating forward in time, the neutral variants carried by
individuals have no effect on the dynamics of the simulation.  \forqs takes
advantage of this by ignoring (i.e. not storing in memory) any ancestral
neutral variation on the founding haplotypes.  Each new mutation generated in
\forqs results in a new haplotype id.  Because of this, it is necessary to use
an id ancestry map to translate the new ids back to the ancestral ids before
performing the mapping.  After the neutral variation has been mapped, the new
mutations can then be merged with the ancestral variants.
This procedure is detailed in the following shell script (and data files
referenced in the script), which runs a complete example of a \forqs simulation
followed by mapping \texttt{ms} sequences (with new mutations merged):
\begin{small}
\begin{verbatim}
    examples/ms_map_example_new_mutations.sh
\end{verbatim}
\end{small}

\newpage

\section{Validation}

\forqs has an extensive set of unit tests that verify the correctness of
individual code modules.  In addition, in order to validate the larger-scale
behavior of the simulations, we have compared \forqs simulation results to
theoretical predictions from population genetics and quantitative genetics.

\subsection{Single locus selection}

We first considered simple scenarios where an individual's fitness is determined by 
that individual's genotype at a single locus, under a wide range of fitness effect
sizes and dominance values.  We compared the simulated allele frequency trajectories
to the deterministic trajectories predicted in the limit of infinite population size.

\begin{figure}[H]
    \begin{center}
        \includegraphics[width=.49\textwidth]{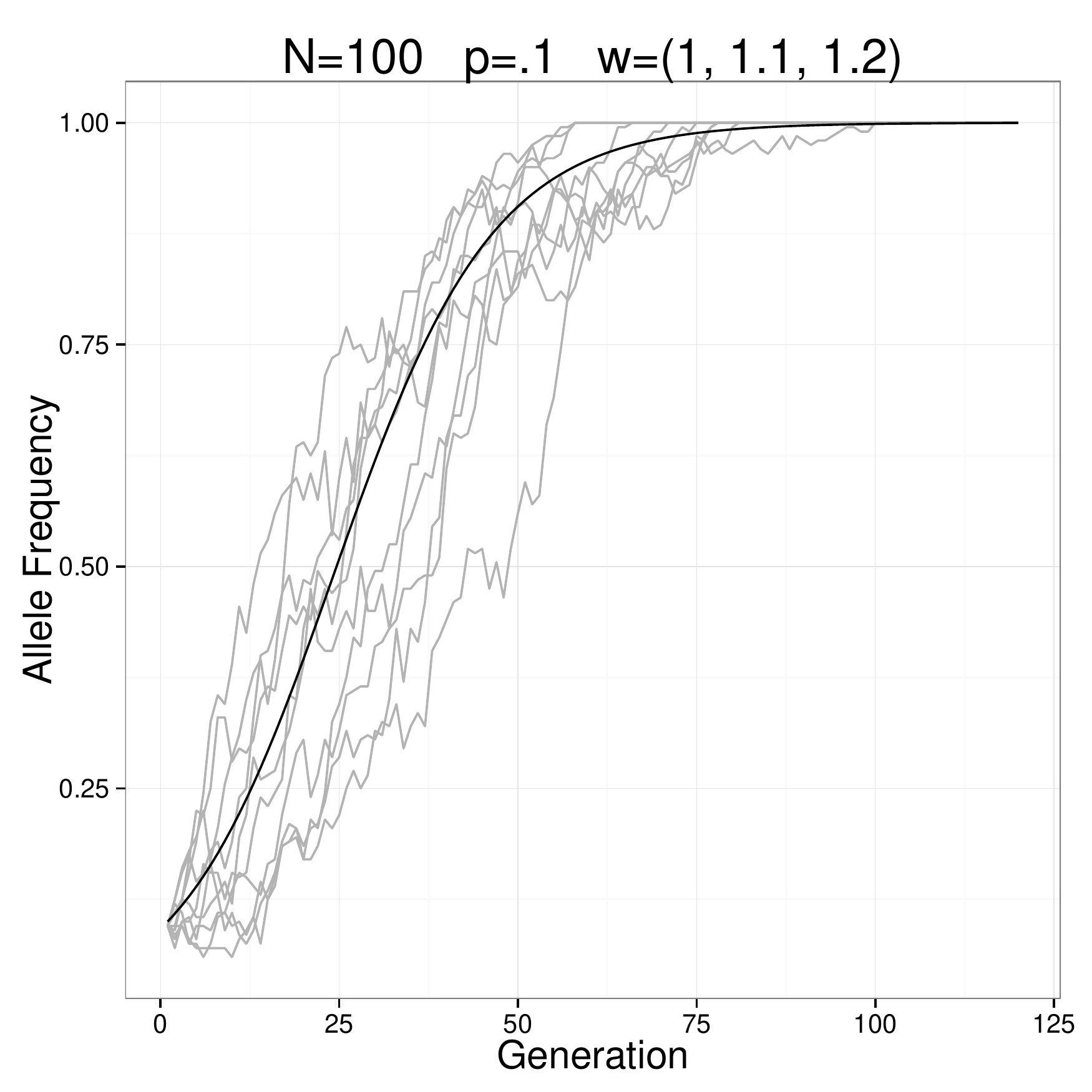}
        \includegraphics[width=.49\textwidth]{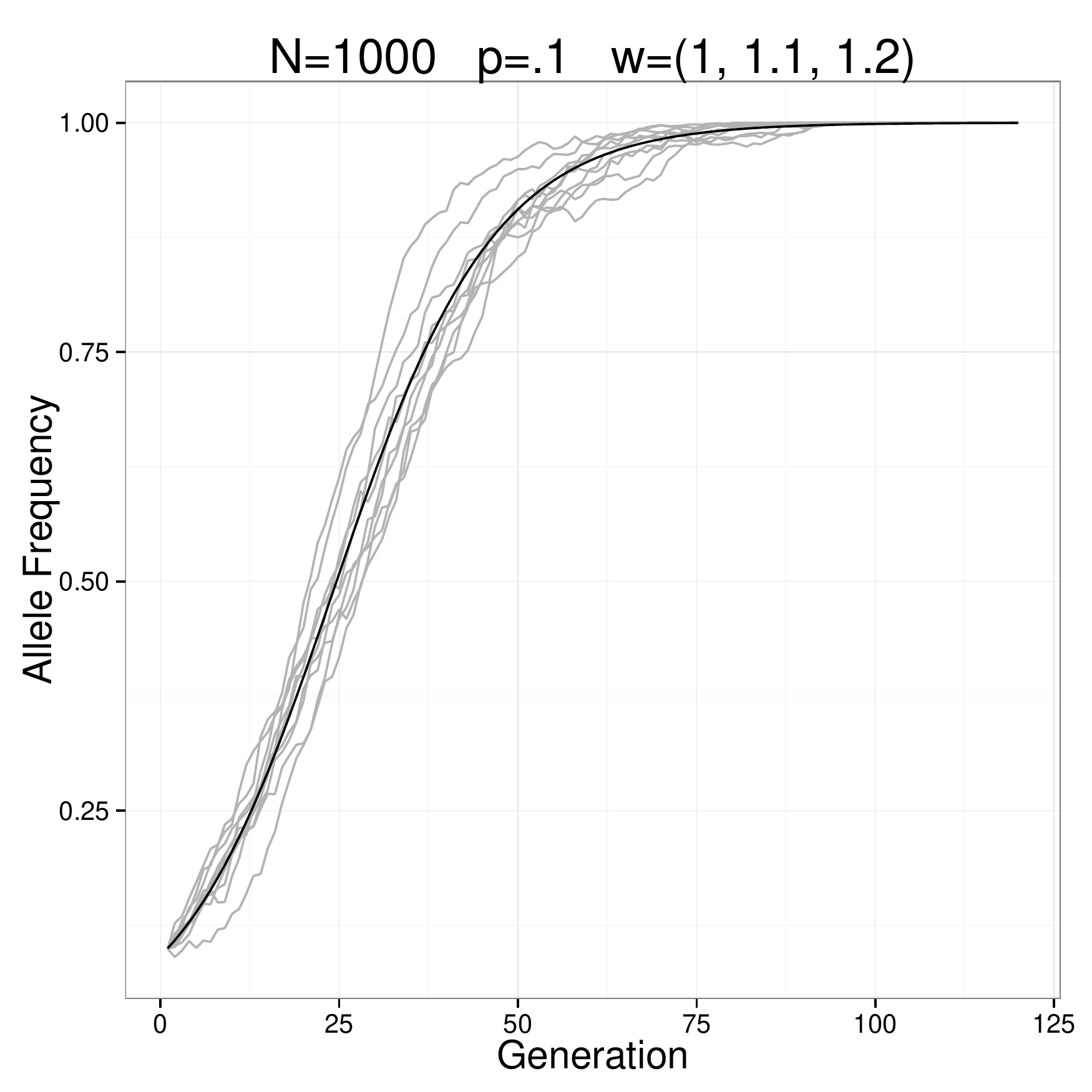}
    \end{center}
    \caption{{\bf Selection on a single locus}.  The dark curves show the deterministic
             trajectories, and the lighter curves show simulated trajectories.  
             Populations of size 100 and 1000 were simulated with additive fitness effect .1 at a single locus,
             with the selected variant having initial allele frequency .1.}
    \label{figure_validation_selection}
\end{figure}

In all cases, the simulated trajectories closely followed the deterministic trajectories,
with better agreement in simulations with larger population sizes, as expected.  Figure
\ref{figure_validation_selection} shows the case of additive positive selection, for
population sizes of 100 and 1000.

\subsection{Decay of linkage disequilibrium}

\begin{figure}[H]
    \begin{center}
        \includegraphics[width=.49\textwidth]{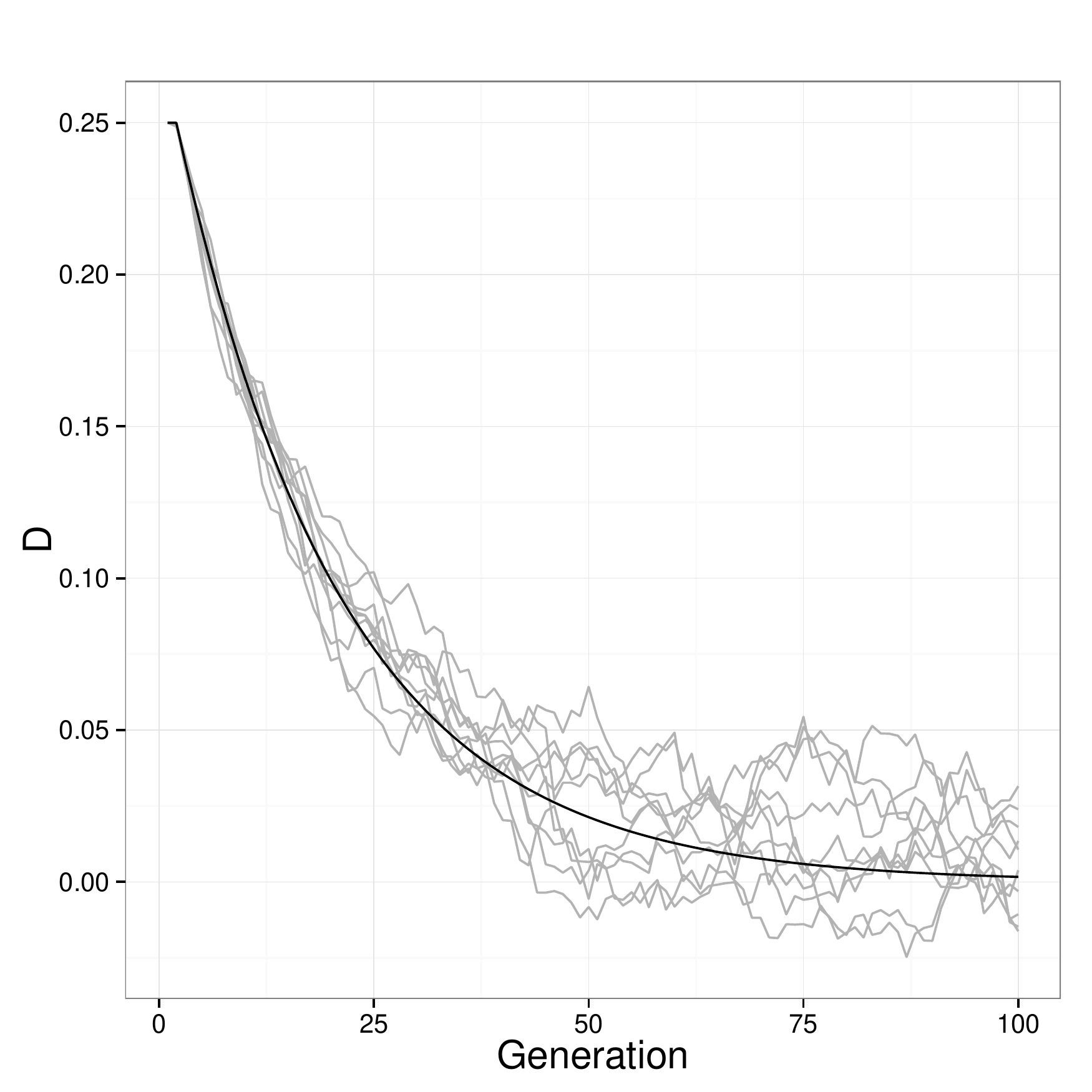}
    \end{center}
    \caption{{\bf Decay of linkage disequilibrium}.  The dark curve shows the deterministic
             trajectory, and the lighter curves show simulated trajectories.  In this example,
             $D=.25$ initially and $r=.05$.
             }
    \label{figure_validation_ld_decay}
\end{figure}

In the limit of infinite population size, linkage disequilibrium (measured by $D$, the correlation
between two variants at different sites) decays geometrically at rate $1-r$, where $r$ is the
recombination rate (see Figure \ref{figure_validation_ld_decay} for an example run).

\subsection{Mutation-drift equilibrium}

To validate our mutation implementation, we approximated the infinite-sites
model by simulating mutations generated randomly in large regions (1mb).  We
set the per-site mutation rate to make the population-scaled mutation rate
$\theta = 10$, and simulated for enough generations to reach mutation-drift
equilibrium.  

We then compared the resulting site-frequency spectra $\{\xi_i\}$ to
theoretical predictions, where $\xi_1$ is the number of singletons, $\xi_2$ is
the number of doubletons, etc. in the sample.  Coalescent theory predicts that
$\xi_i = \theta/i$ in expectation \citep{Fu1995}, under the assumption that the
sample size is small compared to the population size.  When the sample size is
close to the population size, the Wright-Fisher model is expected to have
approximately 12\% more singletons and 2\% less doubletons compared to the
coalescent expected values \citep{WakeleyTakahashi2003}.

\begin{figure}[H]
    \begin{center}
        \includegraphics[width=.49\textwidth]{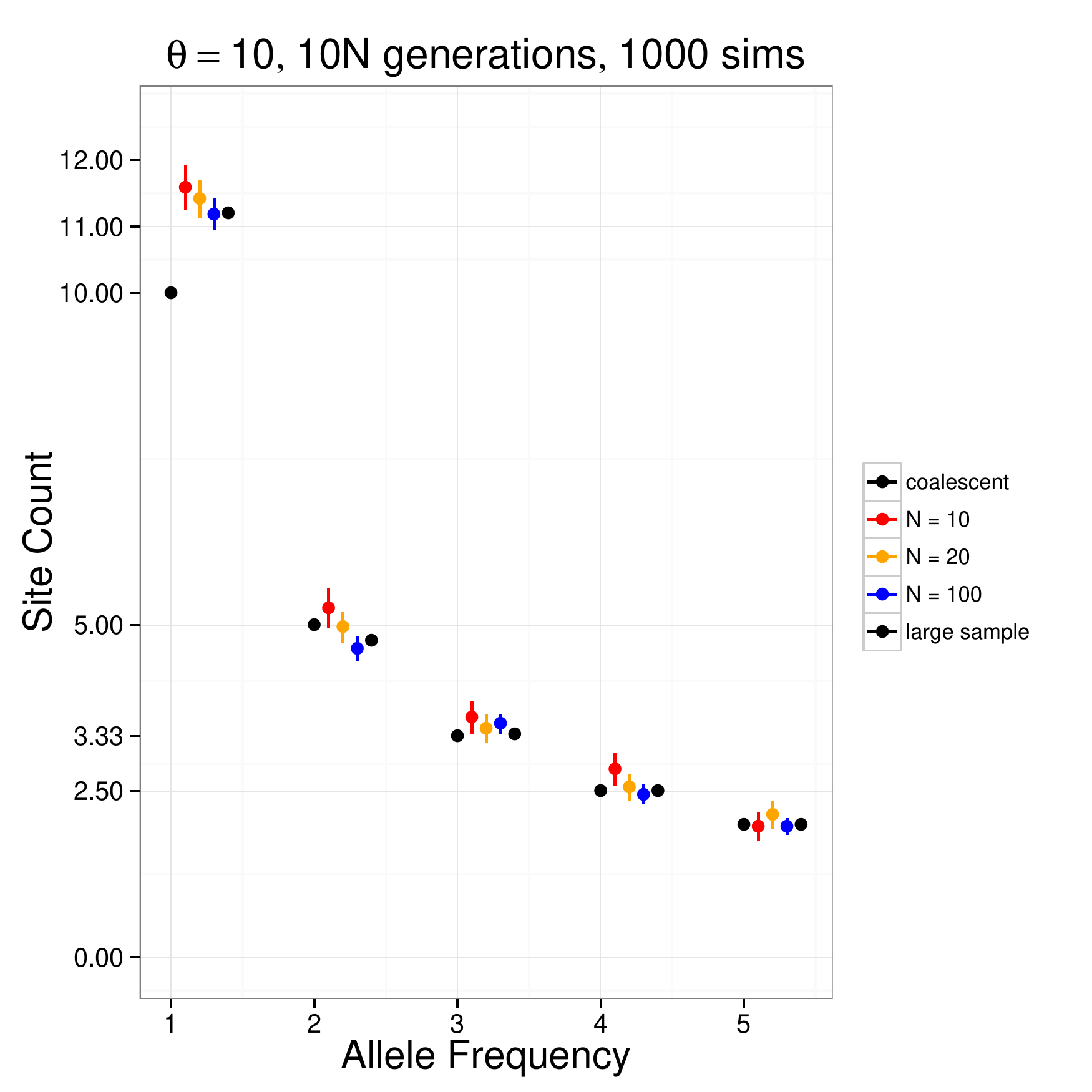}
        \includegraphics[width=.49\textwidth]{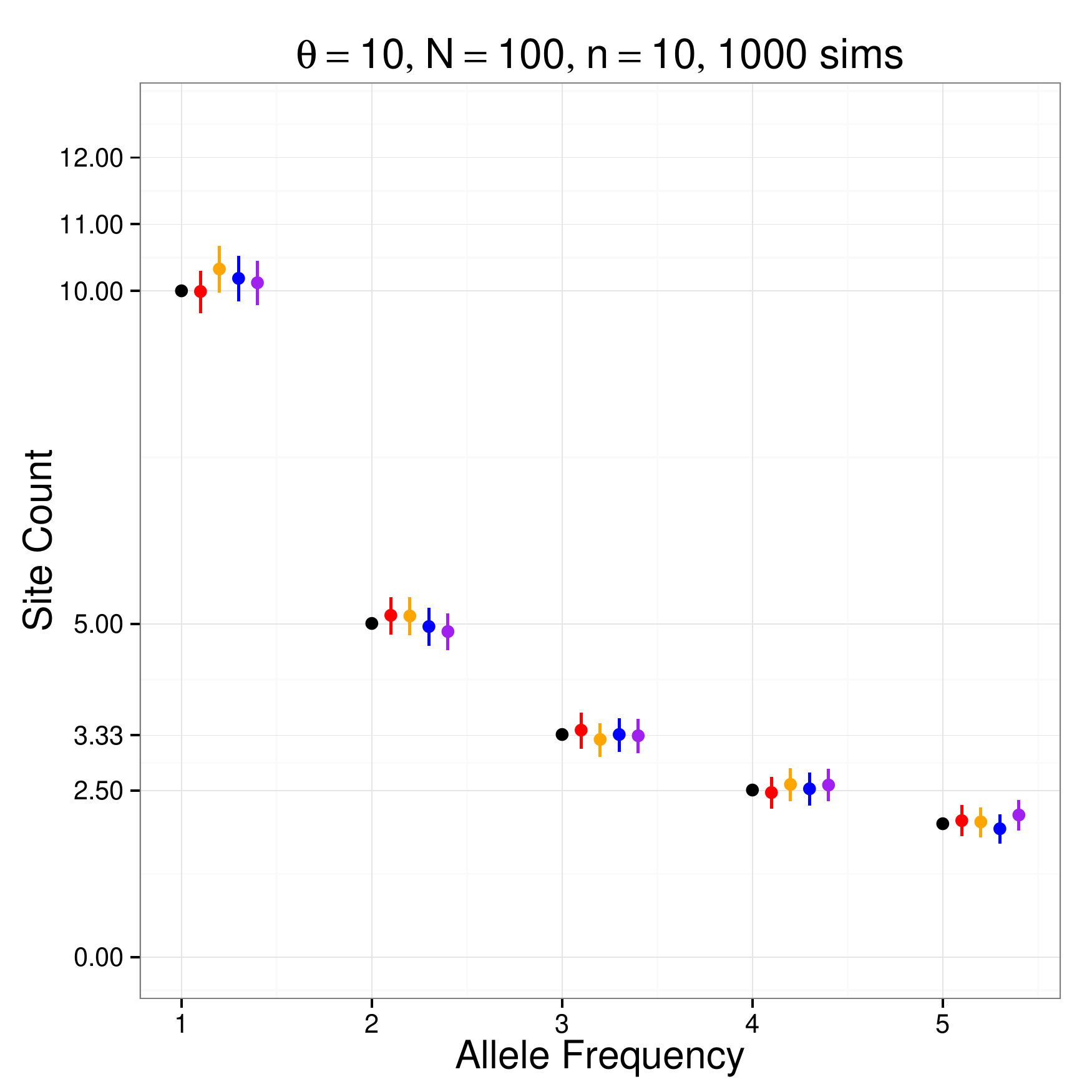}
    \end{center}
    \caption{{\bf Mutation-drift equilibrium}.  
            (\emph{left}) Simulations for different population sizes are shown, together
        with expected values for both the coalescent and Wright-Fisher with large sample size.
        (\emph{right}) Small samples taken from a larger simulated population agree with
        the coalescent expected values.}
    \label{figure_validation_mutation_drift}
\end{figure}

Site frequency spectra generated from the full population agree with the
Wright-Fisher large sample expected values, and site frequency spectra from
smaller samples agree with the coalescent expected values, as expected (Figure
\ref{figure_validation_mutation_drift}).

\subsection{Response to selection}

To validate our implementation of selection on quantitative traits, we
simulated a quantitative trait with 10 QTLs with identical additive effect
sizes and initial allele frequencies, which gives an approximately normal
distribution of trait values in the population.  We also used the \forqs
fitness function \texttt{FitnessFunction\_TruncationSelection}, which selects a
specified proportion of individuals at the upper tail of the trait value
distribution to produce offspring for the next generation. 

The Breeder's Equation from quantitative genetics \citep{GillespieBook,
FalconerMackayBook} predicts the response to selection $R$ from the
heritability of the trait $h^2$ and the selection differential $S$:
\begin{align*}
    R &= h^2 S
\end{align*}
where the $S$ and $R$ are the selected parent mean and offspring mean,
respectively, measured as deviations from the population mean.

\begin{figure}[H]
    \begin{center}
        \includegraphics[width=.9\textwidth]{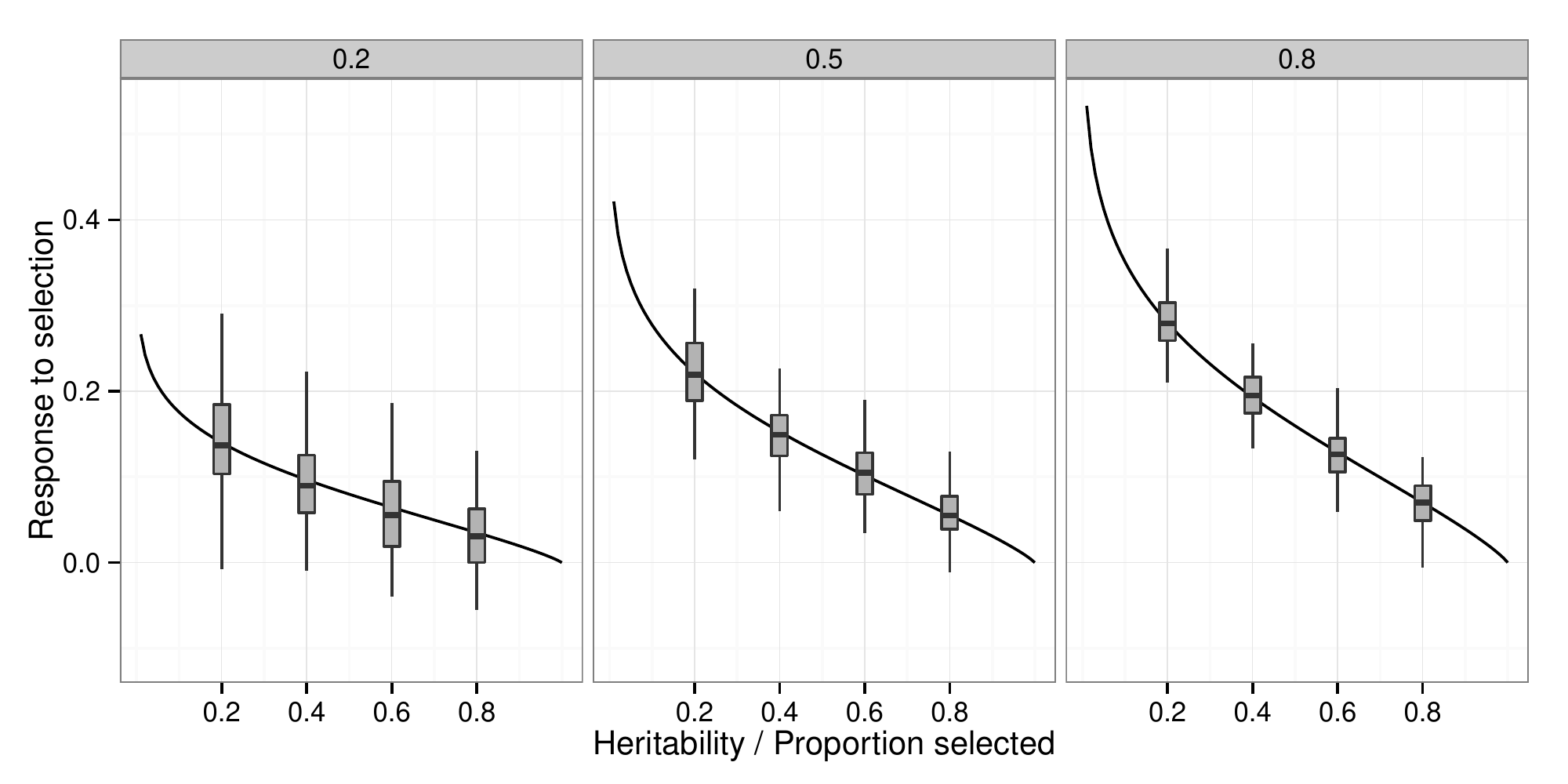}
    \end{center}
    \caption{{\bf Response to selection} 3 panels show simulations run at different
    heritability levels (.2, .5, .8), with varying proportions of selected individuals
    (.2, .4, .6, .8) (100 simulations each).  
    The dark curves show the response values predicted from the Breeder's Equation.
            }
    \label{figure_validation_selection_response}
\end{figure}

Simulations run with various values for the heritability and proportion of
individuals selected agree with the response values predicted from the Breeder's
Equation (Figure \ref{figure_validation_selection_response}).

\newpage

\section{Software development}

\subsection{Building the program}

\forqs is written in C++, and makes extensive use of the Boost libraries, including
the Boost build system.  It is regularly built and tested on OSX and Linux.  In addition,
Windows binaries are built using cross-compilation with gcc on Linux.

The code can be obtained via git from bitbucket:
\begin{small}
\begin{verbatim}
    git clone https://bitbucket.org/dkessner/forqs.git
\end{verbatim}
\end{small}

If you have the both the Boost libraries and the Boost build system installed,
you can build \forqs with the \texttt{bjam} tool, which is the Boost build system's 
equivalent of \texttt{make}:
\begin{small}
\begin{verbatim}
    cd forqs/src
    bjam
\end{verbatim}
\end{small}

\subsection{Structure of the codebase}

Most code units in the \forqs project consist of three files:
\begin{itemize}
    \item interface (header file)
    \item implementation (cpp file)
    \item unit test (cpp file)
\end{itemize}

For example, the code for the \texttt{Locus} module can be found in the files
\texttt{Locus.hpp}, \texttt{Locus.cpp}, \texttt{LocusTest.cpp}.

The unit test for each code unit is run during the build; the build is
successful only if all unit tests pass.

In addition to the unit tests, there are also regression tests, consisting of
scripts in the \texttt{regression\_test} directory.  These tests prevent
unintended behavior changes in the software during new development.  Most of
the tests are based on the example configuration files in the
\texttt{examples} directory, and consist of running \forqs and performing a
\texttt{diff} comparison with known output.  The regression tests are
controlled by a Makefile, and run by \texttt{make}.

The codebase includes all documentation, including the Latex source code for 
this document.  The \forqs Module Reference is generated by Doxygen from
the source code -- each module is documented using Doxygen markup in the
module's header file, immediately preceding the module's class declaration.

\subsection{Software architecture}

This section describes some of the design details of \forqs that may be of
interest to programmers who would like to add new features.

\subsubsection{Low-level data structures}

These are the basic low-level data structures used by \texttt{forqs}:
\begin{itemize}
    \item \texttt{HaplotypeChunk}: a pair of integers (\emph{position}, \emph{id})
    \item \texttt{Chromosome}: an array of \texttt{HaplotypeChunks}
    \item \texttt{Organism}: an array of pairs of \texttt{Chromosomes}
    \item \texttt{Population}: an array \texttt{Organisms}
\end{itemize}

In this setup, \texttt{Population} can be seen as an array of arrays, which
requires a significant number of memory allocations.  To avoid these memory
allocations, \texttt{Population} was re-implemented as a two dimensional
array of \texttt{Chromosome} pairs, resulting in a roughly 30\% speedup.

\begin{sloppypar}
The implementations produce identical results, and can be used interchangeably.
To accomplish this, \texttt{Population} is actually an interface class,
with two concrete implementations: \texttt{Population\_Organisms} and 
\texttt{Population\_ChromosomePairs}.  While the latter is faster, the former
can be more convenient for testing purposes.
\end{sloppypar}

\begin{sloppypar}
In order to handle both cases in a unified manner, the class
\texttt{ChromosomePairRange} represents the begin/end of chromosome pairs
belonging to a single individual, and \texttt{ChromosomePairRangeIterator}
does the appropriate iteration through individuals, depending on the
memory layout of the population.
\end{sloppypar}

\subsubsection{Configurable modules}

\begin{sloppypar}
Configurable modules in \forqs are represented by the \texttt{Configurable}
abstract base class, which defines the common interface through which objects
are instantiated and initialized based on parameters specified by the user.
The top-level module (\texttt{SimulatorConfig}), all primary modules 
(e.g. \texttt{PopulationConfigGenerator}, \texttt{Reporter}, etc.), and
all building block modules (e.g. \texttt{Trajectory}, \texttt{Locus}) 
implement the \texttt{Configurable} interface.
\end{sloppypar}

The \texttt{Configurable} interface describes the functionality
required to serialize an object's configuration, which must be representable as
a list of name-value pairs.  (This is not actually very restrictive, since the
value can be anything representable as a string, including an array of
numbers).

\begin{sloppypar}
\texttt{Configurable} objects are instantiated by \texttt{SimulationBuilder\_Generic},
which handles the mapping from the module name to the actual C++ class.  Each
object in the user-specified configuration file is instantiated and registered
by name so that other objects can obtain references to it if necessary.  
\texttt{SimulationBuilder\_Generic} produces the final \texttt{SimulatorConfig}
used by \forqs for the simulation.
(Note on the name: there were other \texttt{SimulationBuilder} classes that 
constructed \texttt{SimulatorConfigs} based on a more limited set of parameters --
these have since been deprecated in favor of the more generic specification
via configuration files.)
\end{sloppypar}

\begin{sloppypar}
After instantiation, \texttt{Configurable} objects are configured/initialized in
two steps.  First, the \texttt{configure()} method sets any parameters specific
to this module that are specified by the user.  Second, after all objects have been 
configured, the \texttt{initialize()} method allows the modules to communicate with
each other.  For example, objects that need information about the lengths of
the chromosomes (e.g. for recombination) can query the 
\texttt{PopulationConfigGenerator}, which has this information.
\end{sloppypar}

\texttt{Configurable} objects are instantiated, configured, and initialized in the 
order they are specified in the configuration file.  It is important to note that
while objects can obtain references to previously instantiated objects during
the \texttt{configure()} step, they should not try to communicate via these
references until the \texttt{initialize()} step, since the objects they refer to
may not be fully initialized until then.

\begin{sloppypar}
The main \texttt{Simulator} object interacts with \texttt{Configurable} objects
through intermediate interfaces.  This allows the main simulation logic to be
separated from the specific behavior implemented by the various
\texttt{Configurable} modules.  For example,
\texttt{Reporter\_AlleleFrequencies} and \texttt{Reporter\_TraitValues} both
implement the \texttt{Reporter} interface.  \texttt{Simulator} will give each
\texttt{Reporter} information about the current populations, but doesn't need
to know what they do with this information.  As another example, the
\texttt{Simulator} gives a \texttt{QuantitativeTrait} object information about
individuals' genotypes and expects to receive trait values for each individual,
but doesn't need to know details about how this is accomplished.
\end{sloppypar}

The \texttt{Configurable} interface also facilitates some features that make
\forqs more user-friendly, because it provides a translation layer between the
user and program internals.  For example, while 0-based indexing is used
internally, from the user's perspective chromosomes and populations are
numbered starting with 1.  Also, \texttt{Configurable} modules can support
multiple alternate parametrizations.  For example, linear trajectories may be
parametrized by slope-intercept or by two endpoints, and exponential distributions
may be specified by the mean or the rate, depending on which is
more natural for a particular scenario.

\subsubsection{Mutation handling}

Because \forqs tracks haplotype chunks rather than sequences of variants,
mutation is necessarily more complicated than recombination.  This is because a
new point mutation essentially creates a new haplotype that is identical to the
original except at the mutated site.  In order to accomplish this, the
\texttt{VariantIndicator} must be able to be updated with the new haplotype and
variant value.  In addition, the haplotype's ancestry must be stored, since
that haplotype chunk may contain other variants known by the
\texttt{VariantIndicator}.  This is implemented with the special
\texttt{VariantIndicator\_Mutable} that is not specified by the user.  Instead,
it is a wrapper class that is automatically instantiated when the user
specifies a \texttt{MutationGenerator}.  This wrapper class provides the
functionality for updating variants, but passes calls through to the
user-specified \texttt{VariantIndicator} for non-mutated loci.

\subsection{Appendix: Boost libraries}

The Boost C++ Libraries are an essential extension to the C++ Standard Library.  
In addition to providing functionality missing from the Standard Library, the
Boost \texttt{filesystem} library and build system insulate the programmer from
many platform-specific details.

If you have administrative access to your computer, you can use a package
manager to install the Boost libraries and build system.  If not, you can
install Boost locally in your home directory.  The following are instructions
for doing this on a Unix-like system (e.g. OSX or Linux):

\begin{footnotesize}
\begin{verbatim}
# download latest Boost package, uncompress

# install boost libraries

cd boost_???                       # go into the uncompressed directory
bootstrap.sh --help                # see options
bootstrap.sh --prefix=$HOME/local  # install in ~/local
./b2                               # builds libraries -- get coffee (this can take a while)
./b2 install                       # installs stuff in ~/local/include/boost and ~/local/lib

# install boost build

cd tools/build/v2
./bootstrap.sh
./b2 install --prefix=$HOME/local  # puts bjam in ~/local/bin, boost-build in ~/local/share

# also:
#   put ~/local/bin in your path (to find bjam)
# you might need this in your environment:
#   export BOOST_BUILD_PATH=$HOME/local/share/share/boost-build

# to avoid bjam warning: No toolsets are configured.
# create the file user-config.jam in either ~ (home) or ~/local/share/boost-build
# with the following line (note the space before ; is necessary):
    using gcc ; 

\end{verbatim}
\end{footnotesize}

\subsection{Appendix: Cross-compilation targeting Windows}

\begin{sloppypar}
Cross-compilation of Windows binaries from Linux can be done using tools from
the \texttt{mingw-w64} project (\url{http://mingw-w64.sourceforge.net/}).  On
Debian-based systems (e.g. Ubuntu, Mint), installing package
\texttt{g++-mingw-w64} will install the necessary tools and dependencies.
\end{sloppypar}

\begin{footnotesize}
\begin{verbatim}
# add this line to user-config.jam, which defines the toolset gcc-windows
# (you need to tell Boost build which g++, ar and ranlib to use):
using gcc : windows : i686-w64-mingw32-g++ : <archiver>i686-w64-mingw32-ar 
    <ranlib>i686-w64-mingw32-ranlib ;

#   build the Boost libraries (or at least filesystem and system) for Windows
#   (you may need to fiddle with these command line parameters -- these worked for me),
#   from your Boost source dir:
bjam toolset=gcc-windows --prefix=$HOME/local_win32 threading=multi  
  target-os=windows link=static threadapi=win32 --without-mpi 
  runtime-link=static --without-python -sNO_BZIP2=1 --layout=tagged

# note: LIBRARY_PATH doesn't appear to work for cross-compilation, but you can
# put the Boost library files directly in /usr/i686-w64-mingw32, where the various
# development headers/libs are installed for mingw-w64

# the forqs Jamroot contains a target ``windows'', which will build Windows binaries
# and put them in forqs/bin_windows:
bjam windows


\end{verbatim}
\end{footnotesize}

%
%

\pdfbookmark[1]{References}{references}

\bibliographystyle{natbib}
\bibliography{forqs_preprint}

\end{document}